\documentclass[11pt]{article}
\usepackage{amsmath,amsfonts,latexsym,graphicx,amssymb}
\usepackage[dvips]{epsfig}
\usepackage{amsfonts}
\usepackage{bm}
\date{}

\newcommand{{\Cd}}{{\mathbb{C}^d}}

\def\<{\langle}
\def\>{\rangle}
\newtheorem{theorem}{Theorem}
\newtheorem{lemma}{Lemma}

\numberwithin{equation}{section}

\begin{document}

\title{\bf Generalized Circulant Densities and a Sufficient Condition for Separability}
\author{Dariusz Chru\'sci\'nski\\
{\it Institute of Physics, Nicolaus Copernicus University,}\\
{\it Grudzi\c{a}dzka 5/7, 87--100 Toru\'n, Poland}
\\ \\
Arthur O. Pittenger\\
{\it Department of Mathematics and Statistics,}
\\
{\it University of Maryland, Baltimore County, Baltimore, MD 21250}}

\maketitle

\begin{abstract}
In a series of papers with Kossakowski, the first author has
examined properties of densities for which the \textit{positive}
\textit{partial} \textit{transposition} (PPT) property can be
readily checked. These densities were also investigated from a
different perspective by Baumgartner, Hiesmayr and Narnhofer. In
this paper we show how the support of such densities can be
expressed in terms of lines in a finite geometry and how that same
structure lends itself to checking the necessary PPT condition and
to a novel sufficient condition for separability.
\end{abstract}

\section{Introduction}

Interest in quantum information theory has dramatically increased as
the effectiveness of using quantum entanglement as a resource for
storing and manipulating information has become more apparent. Early
ideas for applications include Shor's and Grover's algorithms (see
\cite{NC} for example) as well as quantum key distribution
\cite{Ekert1}, and those insights have motivated the surge in
theoretical and experimental work during the last fifteen years.
Since much of quantum information theory is modelled in the context
of finite dimensional composite systems, a subject of particular
interest is determining the presence or absence of entanglement in
terms of the structure of the density matrix modelling the system.

If one is dealing with two $d$--level particles, the bipartite
context, one can model the state of the composite system as a
positive semi-definite trace one matrix $\rho $ on the tensor
product product space $C_{d}\otimes C_{d},$ where $C_{d}$ denotes a
$d$--dimensional Hilbert space over the complex numbers. The system
is said to be \textit{separable} if it is in the convex hull of
densities of the form $\tau _{1}\otimes \tau _{2}$, where each $\tau
_{k}$ is a $d$--dimensional density on its respective space. An
early observation by Peres \cite{peres} was that such separable
states have the positive partial transpose PPT property: they remain
separable under the partial transpose defined, for example, by $\tau
_{1}\otimes \tau _{2}$ $\rightarrow $ $\tau _{1}\otimes \tau
_{2}^{t},$ where the superscript denotes the transpose operation.

Since separable systems can be modelled classically, the two
particles are not entangled. Thus the PPT condition is a necessary
condition for separability, and it has been shown that it is also
sufficient in the tensor dimensions $2\otimes 2$ and $2\otimes 3$
\cite{Hor1}. In higher dimensions, the condition is not sufficient,
as illustrated by a number of examples such as those in \cite{Hor2}
and \cite{pitrub3} among others. Nonetheless it has proved to be a
surprisingly useful criterion, and there are a variety of examples
of densities illustrating that fact. In \cite{ppta} (see also
\cite{ppt0}) the authors observed that many of these examples have a
common property that facilitates checking the PPT property. That
same structure was also defined in \cite{Narn1}, \cite{Narn2}, and
\cite{Narn3}, where the goal was to investigate the geometry of a
particular subset of densities using ``discrete phase space'' as a
tool. That discrete structure has been used extensively in other
contexts -- \cite{Fiv}, \cite{Gib}, and \cite{pitrub2} are just a
few of many examples -- and the first part of this paper relates the
structure of this class of densities to the finite geometry of phase
space and to checking the PPT condition.

In the second part of the paper we show how the structure of these
densities permits the development of a useful sufficient condition
for separability. This condition is an offspring of a sufficient
condition for separability that appears in \cite{pitrub3}, for
example, and that serves as a counterpoint to the necessary PPT
condition, since there are separable densities that don't satisfy
it. The third part of the paper consists of a number of examples,
including some of those in \cite{ppta,ppt0}, \cite{Narn1} and
\cite{Narn2}, and illustrates the use of this structural criterion

We should note in passing that the investigation of separability and
entanglement is an active area of research and there and too many
papers to cite. Two survey papers on separability and on
entanglement are \cite{Lewen1} and \cite{Hor3}. The interested
reader is referred to those surveys for an overview of the subject
and also to \cite{pitrub3} for the derivation of the sufficient
condition that motivates our main result.

\section{Circulant densities and the PPT condition}

Motivated by the examples in \cite{ppta}, we establish notation that
expresses the support of circulant densities in terms of lines in the
context of finite geometry. The densities $\rho $ to be considered are $%
d^{2}\times d^{2}$ positive semidefinite, trace one matrices with a
particular pattern of support, and we begin by distinguishing between the
positions where non-zero entries can appear and the entries themselves. Let $%
M$ be a $d^{2}\times d^{2}$ matrix with entries equal to zero or one and let
\textit{support}$(M)$ denote the set of positions of $M$ with entry equal to
one.

Let $\left\{ B\left( j,k\right) :0\leq j,k<d\right\} $ denote the
constituent $d\times d$ submatrices, so that when $d=3,$ for example,
\begin{equation}
M=\left(
\begin{array}{ccc}
B\left( 0,0\right) & B\left( 0,1\right) & B\left( 0,2\right) \\
B\left( 1,0\right) & B\left( 1,1\right) & B\left( 1,2\right) \\
B\left( 2,0\right) & B\left( 2,1\right) & B\left( 2,2\right)
\end{array}
\right)\ .  \label{M3ex}
\end{equation}
For $M$ to provide the locations for the non-zero entries of a density, we
obviously need symmetry, $M=M^{t}$, and that is equivalent to $B^{t}\left(
j,k\right) =B\left( k,j\right) $ for all $j,k$. Entries of $M$ can be
indexed as $M(r,s),$ $0\leq r,s<d^{2}$ or in tensor product notation as $%
M_{j_{1}j_{2}},_{k_{1}k_{2}}$, $0\leq j_{1},j_{2},k_{1},k_{2}<d.$ The
relationship between the two is
\begin{equation}
\left( r,s\right) \leftrightarrow \left(
dj_{1}+j_{2},dk_{1}+k_{2}\right) \ . \label{tens1}
\end{equation}
In particular, we routinely use the fact that such an entry corresponds to
the $\left( j_{2},k_{2}\right) $ entry of $B\left( j_{1},k_{1}\right) $.

As the authors noted implicitly or explicitly in \cite{ppta} and \cite{Narn2}%
, the support of $M$ can be interpreted as lines in a two dimensional
\textit{module}; that is, they are lines in
\begin{equation}
V_{2}(d)=\left\{ \left( x,y\right) :x,y\in Z_{d}\right\}\ ,
\label{V2d}
\end{equation}
where $Z_{d}$ denotes the integers modulo $d.$ (If $d$ is prime, $%
V_{2}\left( d\right) $ is a \textit{vector} \textit{space}.) A typical
example is the class of lines of the form
\begin{equation}
L_{p}=\left\{ \left( x,y\right) :y=x+p,x\in Z_{d}\right\}\ ,
\label{lines1}
\end{equation}
where the addition is ${\rm mod}\ d,$ and one connects such lines
with the matrix examples in \cite{ppta} by orienting the $y$-axis
down. For example, when $d=3$ and $p=1$, $L_{1}$ can be represented
by the ones in
\[
\left(
\begin{array}{ccc}
0 & 0 & 1 \\
1 & 0 & 0 \\
0 & 1 & 0
\end{array}
\right)\ .
\]
We can also include ``vertical'' and ``horizontal'' sets of lines using $%
L_{p}$ to denote $\left\{ \left( x,y\right) :x=p\right\} $ or $\left\{
\left( x,y\right) :y=p\right\} .$

By analogy with the Euclidean plane, two lines are called \textit{parallel}
if they do not intersect. For two lines in the same class, $L_{p}\cap
L_{q}=\phi $ if $p\neq q,$ and it is easy to confirm the following result.

\begin{lemma}
$\left\{ L_{p}:0\leq p<d\right\} $ is a partition of $V_{2}(d)$ by a set of $%
d$ mutually parallel lines.
\end{lemma}

Note that we haven't defined lines using the more general formula $%
ax+by+c=0. $ This is because we are not assuming that $d$ is a prime, and
thus we cannot use the properties of an algebraic field which guarantee
multiplicative inverses of non-zero elements.

With this notation in hand, we can describe the pattern of non-zero entries
in a generalized \textit{circulant} density as developed in \cite{ppta}. Let
$p$ denote a permutation of $Z_{d},$ with the proviso that $p\left( 0\right)
=0.$ Define the non-zero entries of $B_{p}\left( j,k\right) $ by the line
\begin{equation}
L_{p}\left( j,k\right) =\left\{ \left( x+p(j),x+p(k)\right) :x\in
Z_{d}\right\}\ ,   \label{Emat}
\end{equation}
and note that $B_{p}^{t}\left( j,k\right) =B_{p}\left( k,j\right) $ (When $p$
is the identity permutation, we suppress the subscript.) Then $M_{p}$ itself
is defined in terms of Dirac notation and the $L_{p}(j,k)$ as
\begin{equation}
M_{p}=\sum_{j,k}\left[ \sum_{x=0}^{d-1}\left| j\right\rangle
\left\langle k\right| \otimes \left| p\left( j\right)
+x\right\rangle \left\langle p\left( k\right) +x\right| \right]
=\sum_{x}I_{p}\left( x\right)\ , \label{Mdef1}
\end{equation}
where $I_{p}\left( x\right) =\sum_{j,k}\left| j\right\rangle \left\langle
k\right| \otimes \left| p\left( j\right) +x\right\rangle \left\langle
p\left( k\right) +x\right| .$

As an example, let $d=3$ and let $p$ denote the identity permutation. Using $%
x_{k}$ to denote an entry of one when $x=k$ and dots for entries of zero,
the resulting $M$ matrix is
\[
\left(
\begin{array}{ccccccccc}
x_{0} & . & . & . & x_{0} & . & . & . & x_{0} \\
. & x_{1} & . & . & . & x_{1} & x_{1} & . & . \\
. & . & x_{2} & x_{2} & . & . & . & x_{2} & . \\
. & . & x_{2} & x_{2} & . & . & . & x_{2} & . \\
x_{0} & . & . & . & x_{0} & . & . & . & x_{0} \\
. & x_{1} & . & . & . & x_{1} & x_{1} & . & . \\
. & x_{1} & . & . & . & x_{1} & x_{1} & . & . \\
. & . & x_{2} & x_{2} & . & . & . & x_{2} & . \\
x_{0} & . & . & . & x_{0} & . & . & . & x_{0}
\end{array}
\right)\ .
\]
Notice in particular the arrangement of the non-zero entries in the
different $3\times 3$ blocks: in each $B(j,k)$ there is precisely one point
of support associated with each $x_{k}.$

In confirming the properties of a putative density $\rho ,$ it is easy to
check the trace one and the Hermitian conditions. What is harder is checking
that $\rho $ is positive semidefinite. One of the points of the present
discussion is that if $\rho $ has support in support($M_{p}$), then that
difficulty is reduced by the following observation. Representing a $d^{2}$
vector $v$ in Dirac notation as $\sum_{j,k}v_{jk}\left| j\right\rangle
\left| k\right\rangle $, it is easy to confirm that the entries of $\rho $
are partitioned into disjoint sets indexed by $x$ in the expression:
\begin{equation}
\left\langle v\right| \rho \left| v\right\rangle =\sum_{x}\sum_{j,k}%
\overline{v}_{j\left( x+p\left( j\right) \right) }\rho _{j\left(
x+p\left( j\right) \right) ,k\left( x+p\left( k\right) \right)
}v_{k\left( x+p\left( k\right) \right) }\ .  \label{Posdef1}
\end{equation}
It is also easy to check that the components of $v$ are partitioned into
classes indexed by $x,$ and thus one can check that $\rho $ is positive
semidefinite by checking that $\rho $ restricted to each of the $I_{p}\left(
x\right) $ is positive semidefinite. This means one is checking $d$ $d\times
d$ matrices rather than one $d^{2}\times d^{2}$ matrix.

That same feature simplifies checking the PPT property, as was
established in \cite{ppta} and \cite{Narn2}.

\begin{theorem}
Let $p$ denote a permutation of $Z_{d}$ with $p\left( 0\right) =0.$ Let $%
M_{p}$ be defined as in (\ref{Mdef1}). Then $M_{p}$ is the sum of the
disjoint matrices $\left\{ I_{p}\left( x\right) :x\in Z_{d}\right\} $ and
for each $(j,k)$ and each $x$ there is exactly one index in \textit{support}(%
$B_{p}\left( j,k\right) \cap I_{p}\left( x\right) ).$ If $\rho $ is a
density with \textit{support}($\rho $) $\subset $ \textit{support(}$M_{p}$%
\textit{)}, then $\rho $ is {\rm PPT} if and only if $\rho $
restricted to $I_{-p}\left( y\right) $ is positive semidefinite for
each $y.$
\end{theorem}

\textit{Proof}: The third sentence follows from the definitions. For the
last assertion, let $\rho ^{\tau }$ denote the partial transposition of $%
\rho $ so that
\[
\rho ^{\tau }=\sum_{j,k}\sum_{x}\left| j\right\rangle \left|
x+p\left( k\right) \right\rangle \rho _{j\left( x+p\left( j\right)
\right) ,k\left( x+p\left( k\right) \right) }\left\langle k\right|
\left\langle x+p\left( j\right) \right| \ .
\]
Substitute $x=y-p\left( j\right) -p\left( k\right) $ and reorder the
summations:
\[
\rho ^{\tau }=\sum_{y}\sum_{j,k}\left| j\right\rangle \left|
y-p\left( j\right) \right\rangle \rho _{j\left( y-p\left( k\right)
\right) ,k\left( y-p\left( j\right) \right) }\left\langle k\right|
\left\langle x-p\left( k\right) \right|\ .
\]
We see that the entries of $\rho $ are again partitioned into disjoint sets
indexed by $y,$ and the earlier discussion applies, using the permutation $%
-p,$ and completing the proof.\qquad $\Box $

The $9\times 9$ example above illustrates the idea. In tensor indexing, one
of the three $3\times 3$ matrices of a density $\rho $ and one of the $3%
\times 3$ matrices of $\rho ^{\tau }$ that one would check for positive
semidefiteness are

\[
\rho :\text{~}~\left(
\begin{array}{ccc}
\rho _{00,00} & \rho _{00,11} & \rho _{00,22} \\
\rho _{11,00} & \rho _{11,11} & \rho _{11,22} \\
\rho _{22,00} & \rho _{22,11} & \rho _{22,22}
\end{array}
\right) ~\text{~}~\text{~~}\rho ^{\tau }:\left(
\begin{array}{ccc}
\rho _{00,00} & \rho _{02,10} & \rho _{01,20} \\
\rho _{10,02} & \rho _{12,12} & \rho _{12,20} \\
\rho _{20,01} & \rho _{20,12} & \rho _{21,21}
\end{array}
\right) \ .
\]
A useful necessary and sufficient condition to verify $\rho $ is positive
semidefinite is that the principal leading minors are non-negative. We leave
it to the reader to write out the other submatrices for $\rho $ and its
partial transpose$.$

In the preceding discussion we used addition of indices in the context of $%
Z_{d}.$ When $d$ is the power of a prime, we have another alternative, and
that is to use the Galois field $GF(p^{n}).$ For example when $d=4=2^{2}$ we
could have defined the four parallel lines (right-side up) as
\[
\left(
\begin{array}{cccc}
\lambda +1 & \lambda & 1 & 0 \\
\lambda & \lambda +1 & 0 & 1 \\
1 & 0 & \lambda +1 & \lambda \\
0 & 1 & \lambda & \lambda +1
\end{array}
\right)\ ,
\]
where $GF(2^{2})$ is the set $\left\{ 0,1,\lambda ,\lambda +1\right\} $ and
the table above actually defines addition. (Multiplication uses $\lambda
^{2}=\lambda +1$ and the usual properties of a field.). Since our only
algebraic requirement so far is that addition be commutative, we could have
used this different set of scalars in the prime power case. With the
appropriate notation to index rows and columns, this gives an example of a $%
d=4$ pattern that generalizes the circulant notation and that is not
included in the appendix of \cite{ppta}. The proof of Theorem (2.1)
then goes through unchanged.

\section{Generalized spin matrices and separability}

The PPT condition is a necessary condition for separability and is
expressed in terms of the components of a density in the
computational basis. Using a generalization of the Pauli spin
matrices, one can obtain a sufficient condition for separability of
a density in terms of the coefficients of a particular orthogonal
family of unitary matrices. An accessible reference is
\cite{pitrub1}, and since the geometry of the support of the
densities considered in this paper make them amenable to this
approach, we summarize the salient points here. In \cite{Narn1} and
\cite {Narn2} this family of matrices is used to define entangled
projections, and the papers analyze the convex hull of those
projections.

Let $\widehat{S}=\left\{ S_{j,k}:0\leq j,k<d\right\} $ be defined by
\begin{equation}
S_{j,k}=\sum_{m=0}^{d-1}\eta ^{jm}\left| m\right\rangle \left\langle
m+k\right|\ ,  \label{spnmat1}
\end{equation}
where $\eta =e^{2\pi i/d}.$ One can interpret these matrices as discrete
Fourier transforms of the computational basis matrices $\left|
m\right\rangle \left\langle m+k\right| ,$ a feature that manifests itself in
various applications. One always has $S_{00}$ equal to the identity matrix.
When $d=2,$ $S_{10}=\sigma _{z},$ $S_{01}=\sigma _{x},$ and $S_{11}=i\sigma
_{y}$ in terms of the usual Pauli spin matrices. That accounts for also
calling them generalized Pauli matrices, and for historical reasons they are
also known as (discrete) Weyl-Heisenberg matrices or, possibly more
accurately, as the discrete Weyl matrices \cite{Narn1}.

Two useful properties of these matrices are:
\begin{equation}
S_{j,k}S_{u,v}=\eta ^{ku}S_{j+u,k+v},~\text{~}~\text{~~\thinspace
\thinspace }\left( S_{j,k}\right) ^{\dagger }=\eta
^{jk}S_{-j,-k}=S_{j,k}^{-1}\ . \label{spinmat0}
\end{equation}
Using the trace norm as an inner product, $\left\langle A\right. \left|
B\right\rangle =Tr\left( A^{\dagger }B\right) $, it follows that $\widehat{S}
$ is an orthogonal family of unitary matrices. As such, $\widehat{S}$ is a
basis for the space of $d\times d$ matrices, and a density can be written in
this basis as
\begin{equation}
\rho =\frac{1}{d}\left[ \sum_{j,k}s_{j,k}S_{j,k}\right]\ ,
\label{spinmat2}
\end{equation}
where $s_{0,0}=1$ and $s_{j,k}=Tr\left[ S_{j,k}^{\dagger }\rho \right] .$
From earlier work we have the following result.

\begin{theorem}
\cite{pitrub1} If $\rho $ is a bipartite density and $S_{u,v}$ denotes the
tensor product $S_{u_{1},v_{1}}\otimes S_{u_{2},v_{2}}$, then a sufficient
condition for $\rho $ to be separable is that $\sum_{u,v}\left|
s_{u,v}\right| \leq 2,$ where $s_{u,v}$ is the coefficient of $S_{u,v}$ in
the spin representation of $\rho .\qquad \square $
\end{theorem}

This is a relatively strong condition, although it is shown in
\cite{pitrub3} that the condition is sharp for certain families of
densities with $d=2^{n}.$ The relevance here is that we can use the
ideas behind the proof of this theorem and obtain sufficient
conditions for separability for certain families of PPT circulant
densities.

To see the connection, let $p$ denote the identity permutation. Then
substituting $m=u$ , $n=m+t$ and $v=m+k$ we can rewrite $M$ as
\begin{eqnarray*}
M &=&\sum_{u,v}\left[ \sum_{t=0}^{d-1}\left| u\right\rangle \left\langle
v\right| \otimes \left| u+t\right\rangle \left\langle v+t\right| \right]  \\
&=&\sum_{k}\sum_{m}\sum_{n}\left| m\right\rangle \left\langle m+k\right|
\otimes \left| n\right\rangle \left\langle n+k\right|  \\
&=&\sum_{k}S_{0,k}\otimes S_{0,k}\ .
\end{eqnarray*}
Thus the support of $M$ equals the support of the sum of the tensor products
$S_{0,k}\otimes S_{0,k}.$ Since the support of $S_{j,k}$ is the same as the
support of $S_{0,k},$ it is reasonable to assume that the spin matrix
representation for a density with \textit{support}($M$) requires only tensor
products of the form $S_{j_{1},k}\otimes S_{j_{2},k}.$ That is indeed the
case, and we skip the easy verification that the other spin coefficients
equal zero.

When the permutation $p$ is not the identity, a similar analysis can be made
that requires some additional notation, but no new concepts. We defer that
discussion to the Appendix and stay with the assumption that $p$ is the
identity permutation.

The next thing to notice is that the main diagonal of $M$ is disjoint from
any $S_{j_{1},k}\otimes S_{j_{2},k}$ with $k\neq 0.$ Put another way, if $%
\rho _{D}$ denotes the diagonal of a density $\rho $ with support$\left(
M\right) $, then
\begin{equation}
\rho -\rho _{D}=\frac{1}{d^{2}}\sum_{k\neq
0}\sum_{j_{1},j_{2}}s_{j_{1}k,j_{2}k}S_{j_{1},k}\otimes S_{j_{2},k}\
. \label{Rhorep1}
\end{equation}
Our strategy is to express the spin coefficients in terms of the
computational basis entries of $\rho $ and the spin matrices themselves in
terms of projections. In certain cases we can then write $\rho $ as a sum of
tensor products of projections with non-negative coefficients -- and that
satisfies the definition of separability. To be effective we require that $d$
be prime power and for computational simplicity we assume that $d$ itself is
prime.

We provide the calculations in the Appendix and record here the basic
structural result.

\begin{theorem}
Suppose $d$ is prime and support$\left( \rho \right) $ is contained in
support$\left( M\right) $ as defined in (\ref{Mdef1}). Then if $\rho _{D}$
denotes the diagonal of $\rho ,$ there are rank one projections $%
P_{a,1}\left( m\right) $ such that
\begin{equation}
\rho -\rho
_{D}=\sum_{a_{1},a_{2}=0}^{d-1}\sum_{m_{1},m_{2}=0}^{d-1}P_{a_{1},1}\left(
m_{1}\right) \otimes P_{a_{2},1}\left( m_{2}\right) C\left(
a,m\right) \ , \label{sepdecomp1}
\end{equation}
where each $C\left( a,m\right) $ is real and $d^{2}C\left( a,m\right) $
equals
\begin{equation}
\sum_{k\neq 0,n_{1},n_{2}}\eta ^{-\binom{k}{2}\left(
a_{1}+a_{2}\right) -k\left( m_{1}+m_{2}\right) -k\left(
a_{1}n_{1}+a_{2}n_{2}\right) }\rho _{n_{1}n_{2},\left(
n_{1}+k\right) \left( n_{2}+k\right) }\ .  \label{sepdec2}
\end{equation}
\end{theorem}

\section{First application of the structural result}

Equations (\ref{sepdecomp1}) and (\ref{sepdec2}) look a bit daunting, but in
special cases they simplify quite nicely. Specifically, assume that $\rho
_{n_{1}n_{2},\left( n_{1}+k\right) \left( n_{2}+k\right) }$depends only on $%
r=n_{2}-n_{1}$:
\begin{equation}
\rho _{n_{1}\left( n_{1}+r\right) ,\left( n_{1}+k\right) \left(
n_{1}+r+k\right) }=c_{r} \ . \label{rhocond1}
\end{equation}
Then $d^{2}C\left( a,m\right) $ equals
\begin{eqnarray}
&&\sum_{r}c_{r}\sum_{k\neq 0}\eta ^{-\binom{k}{2}\left( a_{1}+a_{2}\right)
-k\left( m_{1}+m_{2}+a_{2}r\right) }\sum_{n_{1}}\eta ^{-n_{1}\left( k\left(
a_{1}+a_{2}\right) \right) }  \label{Ceqn2} \\
&=&d\delta \left( a_{1},-a_{2}\right) \sum_{r}c_{r}\sum_{k\neq 0}\eta
^{-k\left( m_{1}+m_{2}-ra_{1}\right) }  \nonumber \\
&=&d\delta \left( a_{1},-a_{2}\right) \sum_{r}c_{r}\left[ d\delta
\left( m_{1}+m_{2},ra_{1}\right) -1\right] \ .
\end{eqnarray}

\subsection{A special case}

As an example, assume that $c_{r}=0$ for $r\neq 0$:
\[
\rho _{n_{1}n_{2},\left( n_{1}+k\right) \left( n_{2}+k\right)
}=c\delta \left( n_{1},n_{2}\right) \ .
\]
Then
\[
\rho =\rho _{D}+c\sum_{n}\sum_{k\neq 0}\left| n\right\rangle \left|
n\right\rangle \left\langle n+k\right| \left\langle n+k\right|\ ,
\]
so the support of $\rho $ is on the diagonal and on $I(0)$, in the notation
above. (Recall that we have assumed $d$ is prime.) We find that
\[
C\left( a,m\right) =\frac{c}{d}\delta \left( a_{1}+a_{2},0\right)
\left[ d\delta \left( m_{1}+m_{2},0\right) -1\right] \ .
\]
Substituting this in (\ref{sepdecomp1}) gives
\begin{eqnarray}
\rho &=&\rho _{D}-\frac{c}{d}\sum_{a_{1}=0}^{d-1}%
\sum_{m_{1},m_{2}}^{d-1}P_{a_{1},1}\left( m_{1}\right) \otimes
P_{-a_{1},1}\left( m_{2}\right)  \label{Appl2a} \\
&&+c\sum_{a_{1}=0}^{d-1}\sum_{m_{1}=0}^{d-1}P_{a_{1},1}\left(
m_{1}\right) \otimes P_{-a_{1},1}\left( -m_{1}\right) \ . \nonumber
\end{eqnarray}

If $c>0$, we can rewrite this using (\ref{spinmat4}) to obtain
\begin{equation}
\rho =\rho _{D}-cI_{d}\otimes
I_{d}+c\sum_{a_{1}=0}^{d-1}\sum_{m_{1}=0}^{d-1}P_{a_{1},1}\left(
m_{1}\right) \otimes P_{-a_{1},1}\left( -m_{1}\right) \ .
\label{Cpos}
\end{equation}
Thus the density $\rho $ will be separable provided the smallest entry in $%
\rho _{D}$ is at least $c.$

Now suppose that $c<0.$ Then we can rework (\ref{Appl2a}) to obtain
\begin{equation}
\rho =\rho _{D}-\left| c\right| \left( d-1\right) I_{d}\otimes
I_{d}+\left| c\right| \sum_{a_{1}}\sum_{m_{1}\neq
-m_{2}}P_{a_{1},1}\left( m_{1}\right) \otimes P_{-a_{1},1}\left(
m_{2}\right) \ ,  \label{Cneg1}
\end{equation}
and $\rho $ is separable if the smallest entry in $\rho _{D}$ is larger than
$\left| c\right| \left( d-1\right) $.

\subsubsection{Isotropic state (example 2 from \protect\cite{ppt0})}

In this example $c=\lambda /d$ and the diagonal entries are either
$\lambda /d+\left( 1-\lambda \right) /d^{2}$ or $\left( 1-\lambda
\right) /d^{2}.$ Then a sufficient condition for separability is
$\lambda \leq 1/\left( 1+d\right) ,$ which also happens to be the
PPT condition.

\subsubsection{Werner density (example 1 from \protect\cite{ppt0})}

Suppose one starts with a density of the form
\[
\rho =\sum_{j,k}b_{j,k}\left| jk\right\rangle \left\langle kj\right| +\sum_{j%
\neq k}c_{jk}\left| jk\right\rangle \left\langle jk\right|\ ,
\]
so that the second expression on the right has only entries on the diagonal
and the first term on the right includes all of the non-zero off-diagonal
entries. For separability, it would suffice to prove either $\rho $ or its
partial transpose $\rho ^{\tau }$ is separable, and it is convenient to work
with the latter. Define
\[
x_{\pm }=\frac{1}{d}\left[ \frac{1-p}{d+1}\pm \frac{p}{d-1}\right]\
.
\]
The assumption in \cite{ppt0} is that for $j\neq k,$ $b_{j,k}=$ $x_{-}$, and
$b_{jj}=x_{-}+x_{+}.$ The diagonal entries $c_{jk}=x_{+}.$ If $x_{-}\geq 0,$
the sufficient condition trivially holds and the density is separable. If $%
x_{-}<0,$ the sufficient condition for separability from
(\ref{Cneg1}) leads to $p\leq 1/2,$ which again is the PPT
condition.

\subsubsection{DiVincenzo et al example \cite{Diven} (example 3 from \protect\cite{ppt0})}

The off-diagonal terms of $\rho $ are $\left( \left( c-b\right) /2\right)
\left| jk\right\rangle \left\langle kj\right| $ in this example, while the
diagonal terms are either $\left( b+c\right) /2$ or $a=1/d-\left( b+c\right)
\left( d-1\right) /2.$ (There is a misprint for the latter value in \cite
{ppt0}.) Again, the notation suggests working with $\rho ^{\tau }$ to put it
into the notational context of this section. If $b<c,$ we find the
sufficient conditions for separability are q
\[
0\leq b~~~~~~~~{\rm and}~~~~~~~cd^{2}+bd\left( d-2\right) \leq 2\ .~
\]
If $c<b,$ one obtains $0\leq c$ and $bd\left( d-1\right) \leq 1.$ It
is easy to check that these conditions are equivalent, so that $b$
and $c$ are non-negative in both cases and both of the other two
conditions hold. As shown in \cite{ppt0}, those were also the PPT
conditions, so the separability analysis gives a stronger result
than the PPT condition.

\subsubsection{Horodecki example (example 4 from \protect\cite{ppt0})}

In this example, $d=3$ and once again $n_{1}=n_{2}$ in the notation
above is a necessary condition for the support of $\rho $. $c=2/21$,
and the diagonal entries are either $\alpha /21$ or $\left( 5-\alpha
\right) /21.$ Then separability is guaranteed by $2\leq \alpha \leq
3$ while the PPT condition allows $2\leq \alpha \leq 4.$ The theory
used here does not resolve the question of separability when
$3<\alpha \leq 4,$ and it is known that in this case the state is
actually entangled.

\subsection{The general case}

We assume the $c_{r}$'s are real and define $s=\sum_{r}c_{r.}$ If we
substitute into (\ref{sepdecomp1}), we can obtain
\begin{equation}
\rho =\rho _{D}-sI_{d}\otimes
I_{d}+\sum_{c}c_{r}\sum_{a}\sum_{m}P_{a,1}\left( m\right) \otimes
P_{-a,1}\left( ar-m\right)\ .  \label{Appl3}
\end{equation}
An easy case is when each of the $c_{r}$'s is non-negative, since that gives
\[
s\leq \min \left( \rho _{nk,nk}\right)\ ,
\]
as a sufficient condition for separability. It is interesting to
note that for $d=3$, if we have $\sqrt{\sum_{k}c_{k}^{2}}\leq \min
\left( \rho _{nk,nk}\right) ,$ then $\rho $ satisfies the PPT
condition and this is a weaker condition than the sufficient
condition for separability.

As another special case, suppose that $c_{0}<0\leq c_{r}$ for $r\neq 0$ and
also that $s=0.$ Then following the usual approach we can write $\rho $ as a
sum of separable projections with non-negative weights and a term $\rho
_{D}-d\left| c_{0}\right| I_{d}\otimes I_{d}$, giving the obvious sufficient
condition.

\subsubsection{Baumgartner, Hiesmayr and Narnhofer: \protect\cite{Narn1},
\protect\cite{Narn2}}

Define $\left| \Omega _{j,k}\right\rangle =S_{j,k}\otimes
I_{d}\sum_{u}\left| u\right\rangle \left| u\right\rangle $ and the
projection $\widetilde{P}_{j,k}=\left| \Omega _{j,k}\right\rangle
\left\langle \Omega _{j,k}\right| .$ The class of densities studied in \cite
{Narn1} and \cite{Narn2} are defined by
\begin{equation}
\rho =\sum_{j,k}c_{j,k}\widetilde{P}_{j,k}\ ,  \label{Narn4a}
\end{equation}
where the constants are non-negative and sum to one. One finds that
\begin{equation}
\rho _{n\left( n+r\right) ,\left( n+k\right) \left( n+r+k\right) }=\frac{1}{d%
}\sum_{j}c_{j,r}\eta ^{-jk}\ .  \label{Narn4}
\end{equation}
There is no dependence on $n,$ and one calculates
\begin{equation}
C\left( a,m\right) =\frac{1}{d^{2}}\delta \left( a_{1},-a_{2}\right)
\sum_{j}\sum_{r}c_{j,r}\left[ d\delta \left(
a_{1}r-j,m_{1}+m_{2}\right) -1\right]\ .  \label{Narn5}
\end{equation}

In recent work on discrete Wigner functions, there is a connection between
lines in phase space and projections -- see for example \cite{Gib} and \cite
{pitrub4} among many others. This motivated one of the examples in \cite
{Narn1}, and the general case is
\[
c_{j,r}=\left\{
\begin{array}{c}
1/d\, ,\,\,\,\,{\rm if}\,\,\ r=sj+t \\
0\, ,\,\,\,\,{\rm otherwise}
\end{array}
\right. ~\ ,
\]
where $s$ and $t$ are fixed. The geometry is that one is defining a line in
the index set, i.e. a discrete phase space. This includes horizontal lines,
and vertical lines have to be treated separately. We then have $C\left(
a,m\right) $ equals
\[
\frac{1}{d^{2}}\delta \left( a_{1},-a_{2}\right)
\sum_{j}\frac{1}{d}\left[ d\delta \left( j\left( a_{1}s-1\right)
,m_{1}+m_{2}-a_{1}t\right) -1\right]\ .
\]
When $a_{1}s\neq 1$, there is exactly one value of $j$ for which the $\delta
$--function equals $1$ and thus $C\left( a,m\right) =0.$ If $s=0,$ this is
true for all $a$ and $m,$ and $\rho $ is diagonal and therefore separable.
In the remaining cases, let $a_{1}=a_{s}\equiv s^{-1}$, so that $C\left(
a,m\right) $ equals
\[
\frac{1}{d^{2}}\delta \left( a_{1},s^{-1}\right) \delta \left(
a_{2},-s^{-1}\right) \left[ -1+\sum_{j}\delta \left(
ts^{-1},m_{1}+m_{2}\right) \right]\ .
\]
We find
\[
\rho =\rho _{D}-\frac{1}{d^{2}}I_{d}\otimes I_{d}+\frac{1}{d}%
\sum_{m}P_{a_{s},1}\left( m\right) \otimes P_{-a_{s},1}\left(
a_{s}t-m\right)\ .
\]
Since the diagonal entries are $\rho _{n\left( n+r\right) ,n\left(
n+r\right) }=\frac{1}{d}\sum c_{j,r}=\frac{1}{d^{2}}$, it follows that these
densities are separable. From properties developed in the Appendix, one can
also check that they are rank $d$ projections, as noted in \cite{Narn1}.

It is known that the normalized identity is contained in an open set
of separable densities. We can observe that property in this context
by considering parameterized segments of the form
\[
\rho \left( t\right) =\frac{1-t}{d^{2}}I_{d^{2}}+t\rho\ ,
\]
where $\rho $ is defined by (\ref{Narn4a}). We will show that if $t\leq
t_{2}=1/\left( 1+d\right) $ then $\rho \left( t\right) $ is separable. That
same bound holds for Werner densities in the bipartite case, and the bound $%
t_{n}=1/\left( 1+d^{n-1}\right) $ works for multipartite Werner densities
\cite{pitrub5}. The obvious conjecture is that $t_{n}$ works for the
multipartite versions of (\ref{Narn4a}).

The proof is an easy application of the structure result. Ignoring terms
with non-negative coefficients, look at the remaining part of $\rho :$%
\[
\rho _{D}-\frac{1}{d^{2}}\sum_{a}\sum_{m_{1},m_{2}}P_{a,1}\left(
m_{1}\right) \otimes P_{-a,1}\left( m_{2}\right) \sum_{j,r}c_{j,r}=\rho _{D}-%
\frac{1}{d}I_{d^{2}}\ .
\]
The usual requirement takes the form
\[
\frac{1-t}{d^{2}}+\frac{t}{d}\sum_{j}c_{j,r}\geq \frac{t}{d}\ ,
\]
so that for all $r$
\[
t\leq 1/\left( 1+d-d\sum_{j}c_{j,r}\right)\ .
\]
This verifies the sufficiency of $t\leq t_{2}$ for separability.

Finally, we illustrate the structural results for one other example from
\cite{Narn1} where $d=3.$ $\alpha $ and $\beta $ are non-negative parameters
and in that notation
\[
\rho =\frac{1-\alpha -\beta }{9}I_{d^{2}}+\alpha
\widetilde{P}_{1,0}+\beta \widetilde{P}_{2,0}\ .
\]
Putting this in the notation above, $t=\alpha +\beta $, $c_{1,0}=\alpha /t$
and $c_{2,0}=\beta /t$. Then the sufficient condition for separability is $%
t=\alpha +\beta \leq \frac{1}{4}.$ Equation(50) in \cite{Narn1}
provides a necessary and sufficient condition for $\rho $ to be PPT,
and it is easy to confirm the condition for separability satisfies
that constraint.

\section{A class of circulant densities with product entries}

In the examples above, the entries of $\rho $ in (\ref{rhocond1}) did not
depend on $n.$ In this section they do, but in a very structured manner:
\[
\rho _{n\left( n+r\right) ,\left( n+k\right) \left( n+r+k\right)
}=x\left( n,r\right) \overline{x}\left( n+k,r\right)\ .
\]
This seems to be a novel set of examples that is amenable to analysis via
the structure results. The key is that
\[
\eta ^{-\binom{k}{2}\left( a_{1}+a_{2}\right) -k\left( m_{1}+m_{2}\right)
-kn\left( a_{1}+a_{2}\right) -kra_{2}}
\]
can be written as
\[
\eta ^{\binom{n}{2}\left( a_{1}+a_{2}\right) +n\left(
m_{1}+m_{2}+ra_{2}\right) }\eta ^{-\binom{n+k}{2}\left(
a_{1}+a_{2}\right) -\left( n+k\right) \left(
m_{1}+m_{2}+ra_{2}\right) }\ .
\]
This meshes well with $x\left( n,r\right) \overline{x}\left( n+k,r\right) $
to give $d^{2}C\left( a,m\right) +\sum_{n}\sum_{r}\left| x\left( n,r\right)
\right| ^{2}$ equal to
\[
\sum_{r}\left| \sum_{n}x\left( n,r\right) \eta ^{\binom{n}{2}\left(
a_{1}+a_{2}\right) +n\left( m_{1}+m_{2}+ra_{2}\right) }\right| ^{2}\
.  \]

Define
\[
\widetilde{A}\left( r,b,t\right) =\left| \frac{1}{d}\sum_{n}x\left(
n,r\right) \eta ^{\binom{n}{2}b+nt}\right| ^{2}
\]
and
\[
Q\left( r,b,t\right)
=\sum_{b=a_{1}+a_{2}}\sum_{t=m_{1}+m_{2}+ra_{2}}P_{a_{1},1}\left(
m_{1}\right) \otimes P_{a_{2},1}\left( m_{2}\right)
\]
so that up to normalization the $Q^{\prime }s$ are separable. Then the
separable representation is
\[
\rho =\rho _{D}-\sum_{n,r}\left| x\left( n,r\right) \right| ^{2}\frac{1}{%
d^{2}}\sum_{a_{1},a_{2}}I_{d}\otimes I_{d}+\sum_{r,b,t}\widetilde{A}\left(
r,b,t\right) Q\left( r,b,t\right)
\]
and it follows that a sufficient condition for separability is
\[
\min_{m,n}\left( \rho _{mn,mn}\right) \geq \sum_{n,r}\left| x\left(
n,r\right) \right| ^{2}\ .
\]

Denoting one of these densities as $\rho _{x}$ we can define the line
segment
\[
\rho _{x}\left( t\right) =\frac{1-t}{d^{2}}I_{d^{2}}+t\rho _{x}
\]
and obtain a sufficient condition for separability
\[
t\left( 1+d^{2}\left( \sum_{n,r}\left| x\left( n,r\right) \right|
^{2}-\min_{m,n}\left( \rho _{mn,mn}\right) \right) \right) \leq 1\ .
\]
We omit the details.

\section{Appendix: technical details of the proof of Theorem 3.2}

To obtain the representation above of $\rho -\rho _{D},$ we first need to
relate projections to the spin matrices, and here is where we restrict the
discussion by assuming that $d$ is a prime. Fix an index $(j,k)$ and define
\begin{equation}
P_{j,k}(r)=\frac{1}{d}\sum_{m}\eta ^{mr}\left( S_{j,k}\right) ^{m}\
. \label{spinmat3}
\end{equation}
Using the properties above, it is straightforward to check that
\[
P_{j,k}(r)P_{j,k}(s)=\delta \left( r,s\right) P_{j,k}(r)\ ,
\]
and that the family $\left\{ P_{j,k}\left( r\right) :0\leq r<d\right\} $ is
an orthogonal family of rank one projections. (As an aside, for the reader
interested in \textit{mutually} \textit{unbiased} \textit{bases,} this is
one way to construct them when the dimension of the space is a prime power.
See \cite{pitrub2}.) Note that one could interpret (\ref{spinmat3}) as
another discrete Fourier transform of matrices, so that one should be able
to compute the spin matrices from the associated projections. In fact the
formula is
\begin{equation}
\left( \eta ^{r}S_{a,1}\right) ^{t}=\sum_{m}\eta ^{-mt}P_{a,1}\left(
m+r\right)\ ,  \label{spinmat4}
\end{equation}
and this is nothing more than the spectral decomposition of $\left( \eta
^{r}S_{a,1}\right) ^{t}$. When $t=0$ the left side is interpreted as the
identity.

Now using (\ref{spinmat0}) it is easy to confirm that $\left( S_{a,1}\right)
^{k}=\eta ^{a\binom{k}{2}}S_{ak,k},$ provided we interpret the binomial
coefficient as $0$ if $k=0$ or $k=1.$ Since $d$ is prime, for $k\neq 0$ and
given $j$ we can define $a=jk^{-1}$ so that
\begin{equation}
S_{j,k}=\eta ^{-a\binom{k}{2}}\sum_{m}\eta ^{-mk}P_{a,1}\left(
m\right)\ , \label{spnmat5}
\end{equation}
and that completes the first step.

For the second step we use the structure of the support of $\rho $ to write
\[
\rho -\rho _{D}=\sum_{m\neq v}\left[ \sum_{t=0}^{d-1}\rho _{m\left(
m+t\right) ,v\left( v+t\right) }\left| m\right\rangle \left\langle v\right|
\otimes \left| m+t\right\rangle \left\langle v+t\right| \right]
\]
\[
=\sum_{k\neq 0}\sum_{n_{1}}\sum_{n_{2}}\rho _{n_{1}n_{2},\left(
n_{1}+k\right) \left( n_{2}+k\right) }\left| n_{1}\right\rangle
\left\langle n_{1}+k\right| \otimes \left| n_{2}\right\rangle
\left\langle n_{2}+k\right| \ .
\]
Comparing this with (\ref{Rhorep1}) we conclude that the constants
$\rho _{n_{1}n_{2},\left( n_{1}+k\right) \left( n_{2}+k\right) }$
are transforms of the $s_{j_{1}k,j_{2}k}\ ,$
\[
\rho _{n_{1}n_{2},\left( n_{1}+k\right) \left( n_{2}+k\right) }=\frac{1}{%
d^{2}}\sum_{j_{1}.j_{2}}\eta ^{j_{1}n_{1}}\eta
^{j_{2}n_{2}}s_{j_{1}k,j_{2}k}\ ,
\]
and thus that
\begin{equation}
s_{j_{1}k,j_{2}k}=\sum_{n_{1}}\sum_{n_{2}}\eta
^{-n_{1}j_{1}-n_{2}j_{2}}\rho _{n_{1}n_{2},\left( n_{1}+k\right)
\left( n_{2}+k\right) }\ .  \label{rhorep2}
\end{equation}
Substituting (\ref{rhorep2}) and (\ref{spinmat4}) into (\ref{Rhorep1}) gives
Theorem 3.2, up to the confirmation that $C(a,m)$ is real. That last fact
follows readily enough by taking the complex conjugate of $C(a,m),$
substituting $-k$ for $k$ and then making some additional notational changes
to obtain $C(a,m).$

Finally, we show that assuming the permutation $p$ is not the
identity only complicates the notation without generalizing the
discussion. Suppose then that the permutation $p$ defining $M_{p}$
is not the identity. Let $\sigma $ denote the inverse permutation
$p^{-1}.$ Make the substitutions $p\left(
u\right) =m,$ $n$ $=m+t$, and $p\left( v\right) =m+k$ in the expression for $%
M_{p}$ to obtain
\begin{eqnarray*}
M_{p} &=&\sum_{t}\sum_{u}\sum_{v}\left| u\right\rangle \left\langle v\right|
\otimes \left| p\left( u\right) +t\right\rangle \left\langle p\left(
v\right) +t\right| \\
&=&\sum_{k}\sum_{m}\left| \sigma \left( m\right) \right\rangle \left\langle
\sigma \left( m+k\right) \right| \sum_{n}\left| n\right\rangle \left\langle
n+k\right| \\
&=&\sum_{k}S_{0,k}\left( \sigma \right) S_{0,k}
\end{eqnarray*}
where $S_{j,k}\left( \sigma \right) =\sum_{m}\eta ^{jm}\left| \sigma
\left( m\right) \right\rangle \left\langle \sigma \left( m+k\right)
\right|\ .$ Introducing the permutation $\sigma $ in the definition
of the spin matrices is equivalent to a relabeling of the
computational basis in the first space and the class
$\widehat{S}\left( \sigma \right) $ should have the same properties
as $\widehat{S}.$ It is easy to check that is indeed the case, and
thus the only difference that one has in (\ref{sepdecomp1}) is that
the projections $P_{a_{1},1}(m_{1})$ should have a $\sigma $ added
to the notation. Thus, for all practical purposes in studying
classes of examples, we can simply assume that $p$ is the identity.

\section*{Acknowledgement} DC research supported in part by the
Polish Ministry of Science and Education Grant 3004/B/H03/2007/33.
AP research supported in part by NSF Grant 0605069.

\end{document}